# Phase Synchrony Component Self-Organization in Brain Computer Interface

Xu Niu, Na Lu*, Huan Luo, and Ruofan Yan

***Abstract*—Phase synchrony information plays a crucial role in analyzing functional brain connectivity and identifying brain activities. A widely adopted feature extraction pipeline, composed of preprocessing, selection of EEG acquisition channels, and phase locking value (PLV) calculation, has achieved success in motor imagery classification (MI). However, this pipeline is manual and reliant on expert knowledge, limiting its convenience and adaptability to different application scenarios. Moreover, most studies have employed mediocre data-independent spatial filters to suppress noise, impeding the exploration of more significant phase synchronization phenomena. To address the issues, we propose the concept of phase synchrony component self-organization, which enables the adaptive learning of data-dependent spatial filters for automating both the preprocessing and channel selection procedures. Based on this concept, the first deep learning end-to-end network is developed, which directly extracts phase synchrony-based features from raw EEG signals and perform classification. The network learns optimal filters during training, which are obtained when the network achieves peak classification results. Extensive experiments have demonstrated that our network outperforms state-of-the-art methods. Remarkably, through the learned optimal filters, significant phase synchronization phenomena can be observed. Specifically, by calculating the PLV between a pair of signals extracted from each sample using two of the learned spatial filters, we have obtained an average PLV exceeding 0.87 across all tongue MI samples. This high PLV indicates a groundbreaking discovery in the synchrony pattern of tongue MI.***

## I. Introduction

Phase synchronization is a ubiquitous phenomenon in the oscillatory processes of the human nervous system, revealing the cooperative interactions between anatomically distinct neural populations. Due to the high time resolution of electroencephalogram (EEG), phase synchrony is often investigated by quantifying the interaction between different EEG channels. Measuring phase synchrony provides insights into the underlying neural circuit dynamics and mechanisms of complex cognitive behaviors [1, 2].

Due to the significant neurophysiology interpretability, phase synchrony-based methods have achieved large success in motor imagery (MI) classification. Gysels et al. [3] first applied phase locking value (PLV) and spectral coherence to classify EEG signals in the field of brain computer interface (BCI). Subsequently, numerous researchers measured both large-scale and local-scale phase synchronies between and within supplementary motor area (SMA) and primary motor area (M1) for MI classification [4-7]. It was widely accepted that phase synchrony-based features and power spectral density mutually contribute valuable information, and thus can be combined to improve BCI performance. Furthermore, Xu et al. [8] utilized phase synchronization information to classify different MI tasks performed by the same limb, demonstrating its potential to overcome the limitations of unnatural control and low control dimensionality in existing BCIs. Consequently, synchrony-based classification methods have gained increasing attention in recent years, with the incorporation of advanced theories such as Riemannian geometry and deep learning (DL) [9, 10].

However, the existing classification methods often require manual selection of EEG acquisition channels and extraction of phase synchrony features between them. This manual feature engineering process, relying on expert knowledge, is time-consuming and lacks flexibility in various application scenarios. Furthermore, in the pre-processing stage, these methods mainly employ data-independent spatial filters such as common average reference (CAR) and Laplacian filters to enhance signal-to-noise ratio (SNR), or utilize ICA solely for removing ocular and myoelectric artifacts.

It is well-known that data-dependent spatial filtering algorithms such as common space pattern (CSP) and independent component analysis (ICA) can project raw EEG signals into a new space, where more significant phase synchronization phenomenon may be observed. Especially, some of the components extracted by ICA are believed to be more closely related to cortical-inspired brain components. However, to our knowledge, no existing research has applied data-dependent spatial filters represented by ICA for extracting the brain components with phase synchrony information (called as PSC in this paper) from raw EEG. This may be due to the fact that the current data-dependent spatial filters fail to outperform data-independent spatial filters. It is likely that more

This work was supported by National Natural Science Foundation of China under Grant U22B2036. (*Corresponding author: Na Lu*)
X. Niu, *N. Lu, H. Luo, and R. Yan are with Systems Engineering Institute, Xi'an Jiaotong University, Xi'an, Shaanxi 710049, China (e-mail: Niu775097397@stu.xjtu.edu.cn; lvna2009@xjtu.edu.cn; yanruofan@stu.xjtu.edu.cn; luohuan123@stu.xjtu.edu.cn).

Mentions of supplemental materials and animal/human rights statements can be included here.

Color versions of one or more of the figures in this article are available online at http://ieeexplore.ieee.org

effective PSCs exist but are currently challenging to obtain. Overall, the progress of phase synchrony-based classification methods has been restricted by the lack of innovative data-dependent spatial filters.

Therefore, we propose utilizing deep learning networks to adaptively learn data-dependent spatial filters, thereby simultaneously overcoming the aforementioned drawbacks. The learned spatial filters can automatically select crucial channels and fuse them to obtain PSCs.

In fact, spatial filters learned by end-to-end DL networks have shown advantages over the traditional algorithm-based spatial filters in recent years. ShallowNet [11] and EEGNet [12] have demonstrated the ability to approximate CSP-based spatial filters, while KFCNet [13] can emulate the functionality of ICA. These end-to-end networks typically comprise filter, power feature extraction, and classification modules. They directly classify raw EEG by optimizing all module parameters using backpropagated gradient of classification errors. The filter module learns optimal spatial filters for maximum-quality power features when peak classification accuracy is reached.

However, existing end-to-end networks struggle to exploit phase synchrony information, making it challenging to learn filters for the extraction of PSCs. The exploitation of phase synchrony information often involves complex number calculations and division operations, which are difficult to converge when implemented using DL networks.

To address the issues, a novel phase-to-amplitude transcoding algorithm (termed as PAT) is developed to indirectly utilize phase synchrony information by transforming it into an amplitude distribution. PAT permits gradient backpropagation, thus can be incorporated into any DL networks. Based on which, a phase synchrony component self-organization network (termed as PSNet) is constructed for automatic PSC extraction and MI classification in an end-to-end manner. "Self-organization" encompasses three aspects: First, PSNet autonomously organizes the search for PSCs from raw EEG signals using a spatial convolution layer. Second, the sub-band components of the PSCs are organized into multiple PSC pairs (termed as PSP). Based on the proposed PAT, a phase-to-amplitude transcoder is embedded into PSNet to extract stable and discriminative phase synchrony-based features from the PSPs. Finally, the extracted features are organized for classification.

PSNet can be seen as an algorithm with fixed computation procedures but with parameters that are automatically adjusted through supervised training. During the training, PSNet evaluates the effectiveness of the extracted PSPs based on real-time classification performance. The optimal PSPs are obtained when peak classification performance is achieved.

Extensive experiments conducted on two public MI datasets have verified that PSNet outperforms state-of-the-art classification methods. This superior performance provides the foundation for identifying effective PSPs. Furthermore, the extracted PSPs were validated based on their PLVs. More remarkably, average PLV above 0.87 was observed for rarely researched non-feedback tongue MI.

In summary, there are three main contributions:
1) Phase-to-amplitude transcoding algorithm is the first method to utilize phase synchrony information by calculation procedures permitting gradient backpropagation. PAT enables DL networks to learn to extract phase synchrony-based features.
2) PSNet is the first MI classification network which extracts phase synchrony-based features from raw EEG signals and implement classification end-to-end. Its effectiveness has been verified by the extensive experiments.
3) The PSPs extracted by PSNet show significant PLVs. The PLVs in certain frequency band vary according to the types of MI. This verification establishes the feasibility of automatically extracting PSCs using deep learning, introducing a novel research approach for phase synchronization.

## II. METHOD

Section A introduced a phase synchrony component self-organization process based on a DL network, which makes use of the proposed phase-to-amplitude transcoder to indirectly utilize phase synchrony information. In Section B, we explain the mathematical mechanism behind this approach, while Section C provides the detailed structure of the transcoder. Additionally, Section D outlines the mechanism for classifying the amplitude representations obtained by the transcoder.

### A. Phase Synchrony Component Self-Organization Network

To achieve phase synchrony component self-organization, it is necessary to extract potential PSPs and evaluate their effectiveness. PSNet is constructed to automate this self-organization process.

As shown in Fig. 1, PSNet mainly consists of three convolution layers: SpatialConv, FIRConv, and ClassConv, along with a phase-to-amplitude transcoder and some dimension adjustment operations. The workflow of PSNet involves sub-band component extraction, component pairing, phase-to-amplitude transcoding, and classification procedures. Potential PSPs are extracted by the sub-band component extraction and component pairing procedures, while their effectiveness is evaluated through classification results.

1) **Potential PSP Extraction**

SpatialConv and FIRConv are applied to extract high-quality sub-band components to form potential PSPs.

In SpatialConv, in analogy to independent component analysis (ICA), raw EEG channels are fused to obtain high-SNR filtered signals. An input sample with $C$ channels and time-length $T$ is fed into PSNet through SpatialConv, which contains $F_1$ kernels serving as spatial filters. The output of SpatialConv is of size $F_1 \times 1 \times T$ and composed of $F_1$ PSCs (phase synchrony brain components). Then, the size of the output is transposed as $1 \times F_1 \times T$ to be fed into FIRConv.

The kernels of FIRConv are set as $F_2$ FIR (finite impact response) band-pass filters. $L$ represents the length of the filters. By performing convolution calculation between a PSC and a FIRConv kernel (FIR band-pass filter), a sub-band component of the PSC can be obtained [13]. Thus, the FIRConv output $\mathbf{S} \in R^{F_2 \times F_1 \times N_c}$ is composed of $F_2 \times F_1$ sub-band components, where $N_c = T - L + 1$.

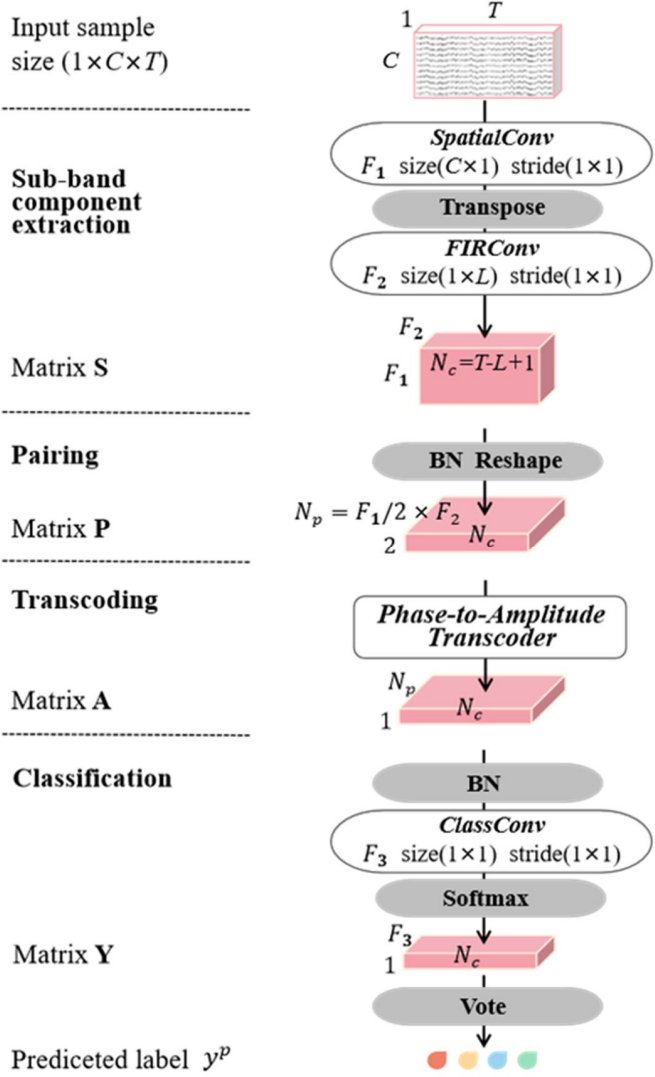

Fig. 1. PSNet structure. Italics denote convolution layers or modules, which are the main blocks of PSNet. Transpose denotes matrix-axis transformation. BN denotes batch normalization.

The pairing procedure reshapes $\mathbf{S}$ to $\mathbf{P}$ with a size of $N_p \times 2 \times N_c$, where $N_p = F_1/2 \times F_2$. The matrix $\mathbf{P}$ contains $N_p$ PSPs, namely $\mathbf{P}_{i,:,:} \in R^{2 \times N_c}$ denotes a PSP. Each PSP is formed by two sub-band components from the same frequency band of every two PSCs.

These extracted PSPs are referred to as potential PSPs, because only a portion of them are true PSPs which can reflect phase synchronization phenomenon.

2) **Effectiveness Evaluation and Network Training**

The proposed phase-to-amplitude transcoder is utilized to transform the phase synchrony information of each potential PSP into amplitude representations. The transcoder serves as the core of PSNet, and its mechanism and structure will be explained in detail in Sections II B and C.

As depicted in Fig. 1, the amplitude matrix $\mathbf{A} \in R^{N_p \times 1 \times N_c}$ contains the amplitude representations of the $N_p$ potential PSPs at $N_c$ sampling time-points, which is generated by the transcoder and then fed into ClassConv. ClassConv gives a predicted soft label to the group of the amplitude representations of each time-point. The ClassConv output $\mathbf{Y} \in R^{F_3 \times 1 \times N_c}$ contains $N_c$ soft labels.

During the training step of PSNet, a cross-entropy value is calculated between each soft label and true label of the input sample. The average of the $N_c$ cross-entropy values serves as the training loss to update network parameters. Notably, FIRConv kernels remains unchanged. SpatialConv is updated to search for high-quality PSCs. The parameters of the transcoder and ClassConv are updated to establish a credible evaluation system for PSP.

During the test step, the final predicted label $y^p$ of the input sample is determined by the majority voting over all the $N_c$ soft labels. The classification accuracy of a test set can be used to evaluate the effectiveness of the PSPs. The optimal self-organization of phase synchrony components is achieved when the highest classification accuracy is attained.

*B. Mathematical Mechanism of Indirectly Utilizing Phase Synchrony Information*

1) **Phase-to-Amplitude Transcoding Algorithm**

Let $s_x(t)$ and $s_y(t)$ represent two narrow-band signals, which from a PSP. $\theta_x(t)$ and $\theta_y(t)$ are their corresponding instantaneous phases. According to [4],

$$\theta_x(t) - \theta_y(t) = \Delta\theta, \quad (1)$$

where $\Delta\theta$ is a constant.

To facilitate understanding, PAT (phase-to-amplitude transcoding algorithm) is initially demonstrated in an ideal case where $s_x(t)$ and $s_y(t)$ have the same invariant amplitude. Therefore, let

$$\begin{cases} s_x(t) = A \sin[\theta_x(t)] \\ s_y(t) = A \sin[\theta_y(t)] \end{cases}. \quad (2)$$

Based on the sum to product formula of trigonometric function, we have

$$\begin{cases} s_x(t) + s_y(t) = 2A \sin\frac{\theta_x(t)+\theta_y(t)}{2} \cos\frac{\theta_x(t)-\theta_y(t)}{2} \\ s_x(t) - s_y(t) = 2A \cos\frac{\theta_x(t)+\theta_y(t)}{2} \sin\frac{\theta_x(t)-\theta_y(t)}{2} \end{cases}. \quad (3)$$

Combining (1) and (3), $A$ can be obtained as

$$A \equiv \sqrt{\left(\frac{s_x(t)+s_y(t)}{2\cos\frac{\Delta\theta}{2}}\right)^2 + \left(\frac{s_x(t)-s_y(t)}{2\sin\frac{\Delta\theta}{2}}\right)^2}. \quad (4)$$

Therefore, the invariant amplitude can be calculated via the sampled values of $s_x(t)$ and $s_y(t)$ at any time. The pair of sampled values is named as a double point couple.

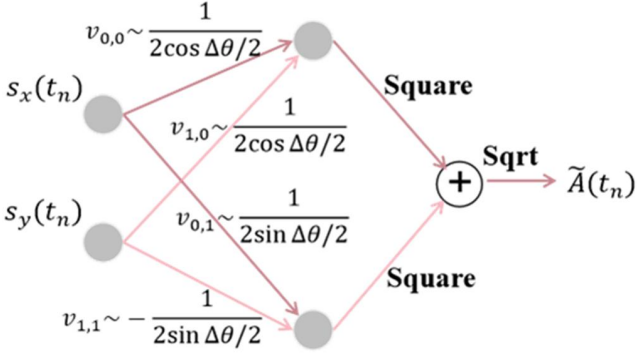

Fig. 2. Mechanism of phase-to-amplitude transcoder. Square denotes quadratic calculation and Sqrt denotes square root calculation.

2) **PSP Judgement Based on Data Distribution**

When $s_x(t)$ and $s_y(t)$ are real-world signals, the double point couples sampled at different times are transformed into different amplitude representations based on (4). The amplitude representation obtained via PAT is called PAT amplitude.

A large number of potential PSP can be extracted from raw EEG by SpatialConv, FIRConv, and the pairing procedure during training PSNet. It is crucial for self-organization that the data distribution of the PAT amplitude of each extracted potential PSP can be used to judge whether it is a true one.

To formally describe the judgement process, let's consider two single-channel EEG signals. One signal has a PSP composed of two sub-band components $s_{1,x}(t)$ and $s_{1,y}(t)$, while $s_{2,x}(t)$ and $s_{2,y}(t)$ extracted from the other signal in the same manner are not a PSP. Two data sets $\{[s_{1,x}(t_n), s_{1,y}(t_n)]\}_{n=1}^{N}$ and $\{[s_{2,x}(t_n), s_{2,y}(t_n)]\}_{n=1}^{N}$ are built by sampling the double point couples of the two signals at time $t_n (n = 1,2,\cdots,N)$. Considering real-world signal,

$$\begin{cases} s_{1/2,x}(t_n) = A_{1/2,x}(t_n)\sin[\theta_{1/2,x}(t_n)] \\ s_{1/2,y}(t_n) = A_{1/2,y}(t_n)\sin[\theta_{1/2,y}(t_n)] \end{cases}, \quad (5)$$

where $\theta_{1,x}(t_n) - \theta_{1,y}(t_n) = \Delta\theta$ but $\theta_{2,x}(t_n)$ and $\theta_{2,y}(t_n)$ are not synchronous. According to [7], we define $[A_{1/2,x/y}(t_n)]_{n=1}^{N}$ obey Rician distributions $R_{1/2,x/y}$.

Based on (4), the PAT amplitude sets $\{\tilde{A}_1(t_n)\}_{n=1}^{N}$ and $\{\tilde{A}_2(t_n)\}_{n=1}^{N}$ are built by

$$\tilde{A}_{1/2}(t_n) = \sqrt{\left(\frac{s_{1/2,x}(t_n)+s_{1/2,y}(t_n)}{2\cos\frac{\Delta\theta}{2}}\right)^2 + \left(\frac{s_{1/2,x}(t_n)-s_{1/2,y}(t_n)}{2\sin\frac{\Delta\theta}{2}}\right)^2}. \quad (6)$$

Since $\theta_{2,x}(t_n)$ and $\theta_{2,y}(t_n)$ are not synchronous, thus estimation value $\tilde{A}_2(t_n)$ is absolutely false. On the contrary, there is only an estimation error between $\tilde{A}_1(t_n)$ and $A_{1,x}(t_n)$ or $A_{1,y}(t_n)$. $\{\tilde{A}_1(t_n)\}_{n=1}^{N}$ obey a data distribution related to $R_{1,x}$ and $R_{1,y}$. Furthermore, an approach to reduce the estimation error between $\tilde{A}_1(t_n)$ and $A_{1,x}(t_n)$ or $A_{1,y}(t_n)$ is proposed in Section II E. Therefore, $\{\tilde{A}_1(t_n)\}_{n=1}^{N}$ and $\{\tilde{A}_2(t_n)\}_{n=1}^{N}$ must obey different data distributions. Based on this distribution difference, a PSP and a non-PSP can be distinguished.

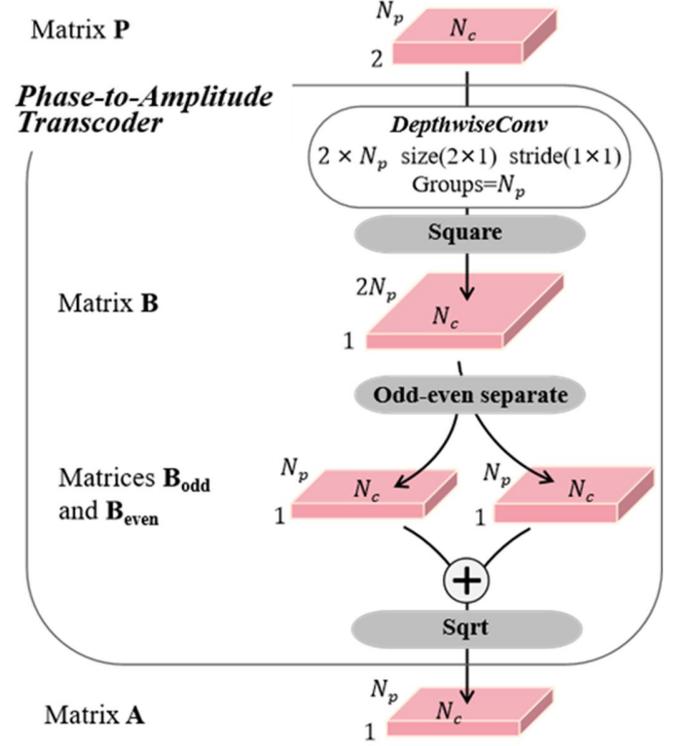

Fig. 3. Structure of the phase-to-amplitude transcoder in PSNet. Square denotes a quadratic activation function and Sqrt denotes a square root activation function.

Accurate classification heavily relies on distinguishing different MI types that exhibit distinct phase synchrony patterns. The key lies in leveraging the data distribution of PAT amplitudes to identify true PSPs. This knowledge enables PSNet to capture and understand the unique phase synchrony patterns associated with each MI type, leading to accurate classification. This will be explained in detail in Section II D.

In summary, while phase synchrony in a pair of sub-band components can be theoretically judged, the unknown value of $\Delta\theta$ and insignificant distribution differences in practical applications may pose challenges. Therefore, in Section II C, deep learning will be introduced to learn the unknown knowledge and information.

*C. Structure of Phase-to-Amplitude Transcoder*

The proposed phase-to-amplitude transcode is designed to imitate the form of (4) but the ability to autonomously learn unknown parameters.

As shown in Fig. 2, the phase-to-amplitude transcoder for a single potential PSP is realized via a full-connected layer. The PAT amplitude of a double point couple $[s_x(t_n), s_y(t_n)]$ sampled at time $t_n$ is obtained as

$$\tilde{A}(t_n) = \sqrt{[v_{0,0}s_x(t_n) + v_{1,0}s_y(t_n)]^2 + [v_{0,1}s_x(t_n) + v_{1,1}s_y(t_n)]^2}. \quad (7)$$

where $v_{0,0}$, $v_{1,0}$, $v_{0,1}$ and $v_{1,1}$ are the weight parameters of the full-connected layer, corresponding to the coefficients of $s_x(t)$ and $s_y(t)$ in (4).

The phase-to-amplitude transcoder used in PSNet is constructed by combing multiple single-PSP phase-to-amplitude transcoders working in parallel. As shown in Fig. 3, the matrix $\mathbf{P}$ consisting of $N_p$ PSPs is fed into the transcoder through a depthwise convolution layer (termed as DepthwiseConv). DepthwiseConv assigns two kernels totaling 4 weight parameters for each PSP in the matrix P. The square activation function is applied after DepthwiseConv. In this way, each double-point couple of each PSP can be transformed to two quadratic terms corresponding to the terms beneath the root sign in (7). These two quadratic terms of all double-point couples form $\mathbf{B} \in R^{2N_p \times 1 \times N_c}$. For each double-point couple, one of its quadratic terms is located in an odd number channel, while the other one is located in the adjacent even channel.

By the odd-even separating operation depicted in Fig. 3, $\mathbf{B}$ is separated as

$$\mathbf{B}_{odd} = \mathbf{B}_{2i-1,:,:} \text{ and } \mathbf{B}_{even} = \mathbf{B}_{2i,:,:}, \quad i = 1,2,\cdots,N_p. \quad (8)$$

Last, the PAT amplitude matrix $\mathbf{A} \in R^{N_p \times 1 \times N_c}$ is obtained as

$$\mathbf{A} = \sqrt{\mathbf{B}_{odd} + \mathbf{B}_{even}}. \quad (9)$$

The phase-to-amplitude transcoder is embedded into PSNet and its weight parameters are updated based on the training loss.

### D. Classification Based on Joint Conditional Probability

As shown in Fig. 1, $\mathbf{P} \in R^{N_p \times 2 \times N_c}$ contains $N_p$ potential PSPs and $\mathbf{A} \in R^{N_p \times N_c}$ (only preserving effective dimensions) contains the PAT amplitudes of the potential PSPs at $N_c$ sampling points. In Section II B, it has been verified that the PAT amplitudes of PSP and non-PSP obey different data distributions.

We can represent the probability density of the element $\mathbf{A}_{i,j}$ in the data distribution of the PAT amplitudes of a PSP as $f_{psp}(\mathbf{A}_{i,j})$, and the probability density of $\mathbf{A}_{i,j}$ in the PAT amplitude distribution of a non-PSP as $f_{non}(\mathbf{A}_{i,j})$. $\boldsymbol{\sigma}_{psp}^i$ and $\boldsymbol{\sigma}_{non}^i$ are respectively the implicit parameter of $f_{psp}(\mathbf{A}_{i,j})$ and $f_{non}(\mathbf{A}_{i,j})$. Based on Bayes formula, a conditional probability is obtained as

$$P(d|\mathbf{A}_{i,j}, \boldsymbol{\sigma}_{psp}^i, \boldsymbol{\sigma}_{non}^i) = \frac{(1-d) \times f_{psp}(\mathbf{A}_{i,j}) + d \times f_{non}(\mathbf{A}_{i,j})}{f_{psp}(\mathbf{A}_{i,j}) + f_{non}(\mathbf{A}_{i,j})}, \quad (10)$$

where $d \in \{0,1\}$. Specifically, $P(d=0|\mathbf{A}_{i,j}, \boldsymbol{\sigma}_{psp}^i, \boldsymbol{\sigma}_{non}^i)$ is the probability that $\mathbf{A}_{i,j}$ comes from the PAT amplitude distributions of the PSP and $P(d=1|\mathbf{A}_{i,j}, \boldsymbol{\sigma}_{psp}^i, \boldsymbol{\sigma}_{non}^i)$ is the probability that $\mathbf{A}_{i,j}$ comes from the non-PSP.

There is a joint conditional probability among the $N_p$ potential PSPs

$$P(\mathbf{d}|\mathbf{A}_{:,j}, \boldsymbol{\sigma}) = \prod_{i=1}^{N_p} P(d_i|\mathbf{A}_{i,j}, \boldsymbol{\sigma}_{psp}^i, \boldsymbol{\sigma}_{non}^i), \quad (11)$$

where $\boldsymbol{\sigma} = [\boldsymbol{\sigma}_{psp}^i, \boldsymbol{\sigma}_{non}^i]_{i=1}^{N_p}$. $\mathbf{d}$ is a vector of length $N_p$ and $\mathbf{d}_i \in \{0,1\}$. $\mathbf{d}_i = 0$ means the $i$-th potential PSPs in $\mathbf{P} \in R^{N_p \times N_c}$ is a true PSP, while $\mathbf{d}_i = 1$ means the $i$-th potential PSPs is a non-PSP. In a word, $\mathbf{d}$ represents the combination of true and false states of the $N_p$ potential PSP.

ClassConv with trainable parameter matrix $\mathbf{W} \in R^{F_3 \times N_p}$ is applied to give a predicted soft label to $\mathbf{A}_{:,j}$. The soft label is obtained as

$$\mathbf{Y}_{:,j} = \mathbf{W} \times \mathbf{A}_{:,j} = [P(c|\mathbf{A}_{:,j}, \mathbf{W})]_{c=1}^{F_3}. \quad (12)$$

The aim of updating $\mathbf{W}$ during PSNet training is to make

$$P(c|\mathbf{A}_{:,j}, \mathbf{W}) = P(c|\mathbf{A}_{:,j}, \boldsymbol{\sigma}) = \frac{\sum_{\mathbf{d}} P(c|\mathbf{d}) P(\mathbf{d}|\mathbf{A}_{:,j}, \boldsymbol{\sigma})}{\sum_{\mathbf{d}} P(\mathbf{d}|\mathbf{A}_{:,j}, \boldsymbol{\sigma})}. \quad (13)$$

### E. PSNet Embedded with Phase Shifter

1) **Shift Phase for Error Suppression**

Combining (5) and (6), we have

$$\begin{cases} \tilde{A}_1^2(t_n) = \left(\frac{s_{1,x}(t_n)+s_{1,y}(t_n)}{2\cos\alpha}\right)^2 + \left(\frac{s_{1,x}(t_n)-s_{1,y}(t_n)}{2\sin\alpha}\right)^2 \\ A_{1,y}^2(t_n) = \left(\frac{g(t_n)s_{1,x}(t_n)+s_{1,y}(t_n)}{2\cos\alpha}\right)^2 + \left(\frac{g(t_n)s_{1,x}(t_n)-s_{1,y}(t_n)}{2\sin\alpha}\right)^2 \\ g[t_n] = \frac{A_{1,y}(t_n)}{A_{1,x}(t_n)}, \quad \alpha = \frac{\Delta\theta}{2} \end{cases}. \quad (14)$$

For the sake of brevity, estimation error $e(t_n)$ is defined as $|\tilde{A}_1^2(t_n) - A_{1,y}^2(t_n)|$. Thus,

$$e(t_n) = \left| \frac{(g^2(t_n)-1)s_{1,x}^2(t_n)}{4}\left(\frac{1}{\cos^2\alpha} + \frac{1}{\sin^2\alpha}\right) \right. \\ \left. +[g(t_n)-1]s_{1,x}(t_n)s_{1,y}(t_n)\left(\frac{1}{\cos\alpha} - \frac{1}{\sin\alpha}\right) \right|. \quad (15)$$

Obviously, $e(t_n)$ is equals to zero if

$$(\cos\alpha - \sin\alpha)\cos\alpha\sin\alpha = G(t_n) = \frac{[g(t_n)+1]s_{1,x}(t_n)}{4s_{1,y}(t_n)}. \quad (16)$$

However, since $G(t_n)$ varies with time, equation (16) cannot always hold. It is feasible to reduce the upper bound of $e(t_n)$, which can be expressed as

$$e(t_n) = \left| \frac{(g^2(t_n)-1)s_{1,x}^2(t_n)}{4}\left(\frac{1}{\cos^2\alpha} + \frac{1}{\sin^2\alpha}\right) \right| \\ + \left| [g(t_n)-1]s_{1,x}(t_n)s_{1,y}(t_n)\left(\frac{1}{\cos\alpha} - \frac{1}{\sin\alpha}\right) \right|. \quad (17)$$

The minimum upper bound can be obtained when $\alpha = \pm\pi/4$ i.e. $\Delta\theta = \pm\pi/2$. Since the original phase difference of a PSP is unknown, it is impossible to manually adjust it to $\pm\pi/2$. However, by embedding a functional module into PSNet, it might be possible for PSNet to learn and automatically adjust the phase difference to the optimal value.

2) **Shift Phase for Error Suppression**

As shown in Fig. 4, a phase shifter based on convolution layer is embedded into PSNet to obtain PSPs with phase differences closed to $\pm\pi/2$. The output of SpatialConv, denoted as $\mathbf{S}^s \in R^{F_1 \times 1 \times T}$, consists of $F_1$ spatial-filtered signals. Half of these signals are fed into the DepthwiseConv of the phase shifter. In the phase shifting process, each signal undergoes zero-padded convolution, which allows it to adjust its phase by moving along the time axis. This phase adjustment does not alter the length of the signal and only affects its temporal alignment. $\mathbf{W}^{SP} \in R^{F_1/2 \times L_{sp}}$ represents the DepthwiseConv kernels. The DepthwiseConv output $\mathbf{S}^p \in R^{F_1 \times 1 \times T}$ is obtained as

$$\begin{cases} \mathbf{S}^p_{2i-1,:,:} = \mathbf{S}^s_{2i-1,:,:} \\ \mathbf{S}^p_{2i,:,:} = \mathbf{W}^{sp}_{i,:} * \mathbf{S}^s_{2i,:,:} \end{cases}, i = 1,2,\ldots,\frac{F_1}{2}. \tag{18}$$

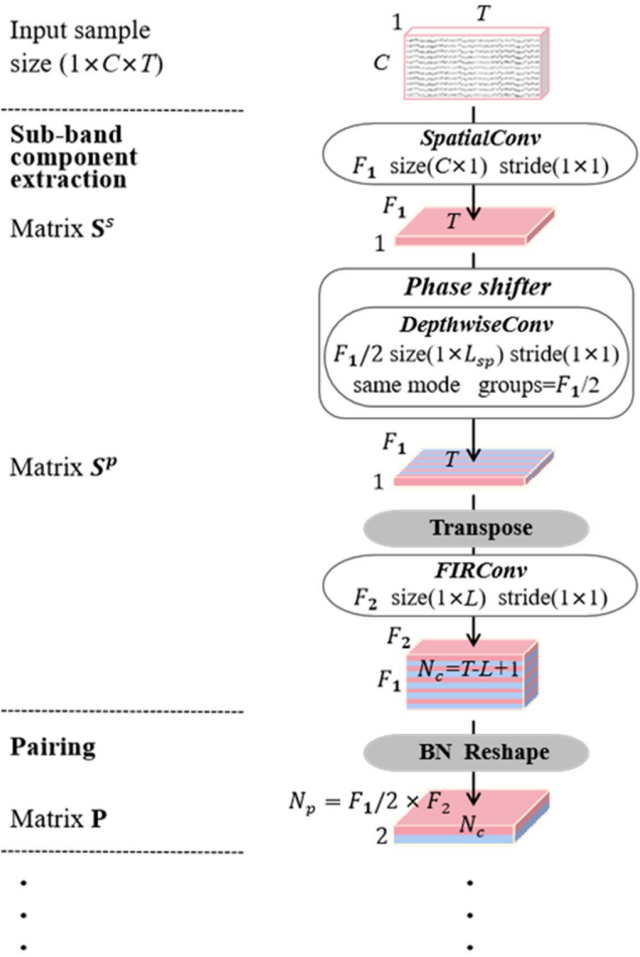

Fig. 4. Phaser-PSNet is constructed by adding a phase shifter between the SpatialConv and transpose operation of PSNet.

The initial $\mathbf{W}^{sp}$ is

$$W^{sp}_{i,j} = \begin{cases} 0, & j \neq (N_s+1)/2 \\ 1, & j = (N_s+1)/2 \end{cases}. \tag{19}$$

Therefore, $\mathbf{S}^p = \mathbf{S}^s$ at the beginning of network training.

Following the same pairing procedures with PSNet, $\mathbf{P} \in R^{N_p \times 2 \times N_c}$ composed of $N_p$ potential PSPs is obtained. Each potential PSP is composed of two sub-band components, which are obtained using two different spatial filters (SpatialConv kernels) and the same FIR band-pass filter (FIRConv kernel).

Different with PSNet, one of the sub-band components in a PSP extracted by phaser-PSNet has moved along time axis.

### III. EXPERIMENT SPECIFICATION

#### A. Sample-Set Description

Experiments were conducted on BCI Competition IV-2a [14] and Motor Movement/Imagery Database (termed as MMIDB) [15, 16].

BCIC IV-2a consists of EEG data from nine subjects with two sessions. Each session for each subject includes 288 trials. The first session was used in our experiments. In each trial, according to an instruction, a subject imagined one of four-category motor imageries (left-hand, right-hand, feet, and tongue). No feedback was provided. The EEG data were collected using 22 electrodes at a sampling rate of 250 Hz. We cropped the EEG signal from 0.5s to 1.5s after the instruction start as a sample. The sample was filtered between 1-48Hz to remove direct-current offset and power line interference. Meanwhile, it was multiplied by 106 to avoid gradient diffusion. Finally, a sample set consisting of 288 samples of size (22, 250) was built for each subject.

MMIDB was recorded via 64 electrodes at a sampling rate of 160 Hz. The first 10 subjects of the 109 subjects were selected for the experiments. Each subject performed 90 EEG trials involving left hand, right hand, both feet, and both hand MI tasks. For each trial, we cropped the EEG signal from 0s to 1s after the start of the instruction as a sample. Similar to BCI Competition IV-2a, we applied a bandpass filter between 1-48 Hz and multiplied the signal by 106 for preprocessing. As a result, a sample set consisting of 90 samples of size (64, 160) was built for each subject.

#### B. Experiment Procedures

Extensive experiments were conducted to verify the performance of PSNet and phaser-PSNet, as well as analyze the self-organized PSPs.

First, PSNet was evaluated with hyper-parameters selected through pre-experiments. Since the focus of this paper is on the feasibility of self-organizing PSPs, effort was not made to find the optimal hyperparameters for maximum performance.

For the experiments on BCIC IV-2a, the hyper-parameters $C$, $F_1$, $F_2$, and $F_3$ were set as 22, 16, 15, and 4 respectively. The length $L$ of the FIR band-pass filters (FIRConv kernels) was set as 51. The center frequency of the FIR band-pass filters was respectively set as $(1 + i \times 2)$Hz where $i = 1, 2, \ldots, 15$. Consequently, the matrix $\mathbf{P}$ extracted by PSNet has a size of $(120 \times 2 \times 200)$, composed of 120 potential PSPs. The Adam optimizer with default parameters and a learning rate of $10^{-3}$ was employed. The training procedure was iterated for 800 epochs. PSNet was evaluated using 4-fold cross-validation, which was repeated 10 times. The highest classification accuracy achieved among the runs was recorded.

For MMIDB, the hyper-parameter $C$ was modified to 64 and $L$ to 33. The remaining hyper-parameters were kept unchanged. PSNet was evaluated by 7-fold cross-validation.

Various state-of-the-art baseline methods have been incorporated for comparison, including ShallowNet [11], ShallowNet with cropped training strategy [11], EEGNet [12], KFCNet [13], 3-D CNN [17], MFBCNN [18], C-LSTM [19], EEG-TCNet [20], and DoNet [21].

TABLE I
PERFORMANCE COMPARISON OF PSNET, PHASER-PSNET AND THE BASELINE METHODS ON BCIC IV 2A (%)

| Method | S1 | S2 | S3 | S4 | S5 | S6 | S7 | S8 | S9 | Ave. |
|---|---|---|---|---|---|---|---|---|---|---|
| C-LSTM | 61.81 | 53.82 | 69.10 | 53.47 | 51.74 | 53.13 | 53.47 | 64.24 | 79.17 | 59.99 |
| MFSBCNN | 62.15 | 51.39 | 60.07 | 67.01 | 56.94 | 51.74 | 61.11 | 59.03 | 77.78 | 60.80 |
| ShallowNet | 63.89 | 54.17 | 71.53 | 52.78 | 51.04 | 55.21 | 56.25 | 61.81 | 82.64 | 61.03 |
| KFCNet | 71.18 | 54.17 | 80.56 | 57.99 | 60.76 | 62.50 | 75.35 | 72.22 | 87.85 | 69.17 |
| ShallowNet (cropped) | 72.22 | 55.56 | 80.56 | 66.67 | 61.46 | 63.19 | 75.00 | 71.87 | 91.32 | 70.87 |
| EEGNet | 76.39 | 65.63 | 76.39 | 73.26 | 71.53 | 61.81 | 84.03 | 72.57 | 87.85 | 74.38 |
| 3-D CNN | 77.40 | 60.14 | 82.93 | 72.29 | 75.84 | 68.99 | 76.04 | 76.86 | 84.67 | 75.02 |
| EEG-TCNet | 73.96 | 56.94 | 84.03 | 71.53 | 71.88 | 69.10 | 78.82 | 80.90 | 89.58 | 75.19 |
| ATCNet | 80.21 | 63.54 | 87.50 | **76.74** | 76.39 | 73.61 | 86.46 | 85.07 | **95.49** | 80.56 |
| PSNet | **85.42** | **67.02** | **89.93** | 68.06 | **76.39** | **73.61** | **92.71** | **89.93** | **95.49** | 82.06 |
| Phaser-PSNet | **87.15** | **67.71** | **92.02** | 71.53 | **78.13** | **74.31** | 92.71 | **90.97** | 95.49 | **83.33** |

TABLE II
PERFORMANCE COMPARISON OF PSNET AND THE BASELINE METHODS ON MMIDB (%)

| Method | S1 | S2 | S3 | S4 | S5 | S6 | S7 | S8 | S9 | S10 | Ave. |
|---|---|---|---|---|---|---|---|---|---|---|---|
| EEG-TCNet | 51.19 | 62.33 | 70.24 | 59.52 | 44.57 | 59.52 | 51.72 | 44.14 | 57.10 | 59.52 | 55.99 |
| MFSBCNN | 53.57 | 70.24 | 66.67 | 57.14 | 53.57 | 70.24 | 57.14 | 51.19 | 60.71 | 59.52 | 60.00 |
| ShallowNet | 54.76 | 75.00 | 69.05 | 60.71 | 60.71 | 73.81 | 61.90 | 59.52 | 67.86 | 60.71 | 64.40 |
| EEGNet | 84.52 | 69.05 | 91.67 | 77.38 | 63.10 | 80.95 | 80.95 | 65.48 | 71.43 | 66.67 | 75.12 |
| ShallowNet (cropped) | 84.52 | 76.19 | 77.38 | 85.71 | 66.67 | 79.76 | 83.33 | 71.43 | 72.62 | 72.62 | 77.02 |
| C-LSTM | 82.14 | 84.52 | 86.90 | 84.52 | 72.62 | 86.90 | 83.33 | 69.05 | 85.71 | **83.33** | 81.90 |
| KFCNet | 80.95 | 84.52 | 91.67 | 88.10 | 72.62 | 86.91 | 91.67 | 82.14 | 83.33 | 79.76 | 84.17 |
| DoNet | 84.52 | **90.48** | 92.86 | 89.29 | 76.19 | **92.86** | **100.0** | 86.90 | **90.48** | 79.76 | 88.33 |
| PSNet | **90.48** | 89.29 | **95.24** | **95.24** | **78.57** | 90.48 | 96.43 | **94.05** | 89.29 | 76.19 | **89.52** |

Second, phaser-PSNet was evaluated in the same manner as PSNet to facilitate a direct comparison between the two models. An extra hype-parameters $L_{sp}$ was set as 51 for BCIC IV-2a.

Lastly, to validate the self-organization process of PSPs in PSNet, the learned PSPs were analyzed using various visualization methods.

## IV. RESULTS AND VISUALIZATION

### A. Performance Comparison

The performance of PSNet, phaser-PSNet, and the baseline methods on BCIC IV-2a is presented in Table I. It can be seen that PSNet outperforms the baseline methods in terms of average accuracy across all the subjects, affirming its effectiveness. The achievement of state-of-the-art performance contributes to advancing research on phase synchrony information mining processes. In more detail, PSNet demonstrates significant superiority over the baseline methods for most subjects, except S4. The baseline methods mainly focus on utilizing power feature for classification. Thus, the relatively weaker performance of PSNet on S4 serves as evidence that PSNet functions based on a distinct mechanism compared to the baselines. Further verification of PSNet's operation based on the phase synchronization phenomenon will be discussed in Section IV B.

By comparing phaser-PSNet and PSNet, it is evident that incorporating the designed phase shifter has resulted in improved accuracy for most subjects. Additionally, phaser-PSNet surpasses PSNet with an average accuracy rate of 83.33% across all subjects. This validates the hypothesis that shifting the phase

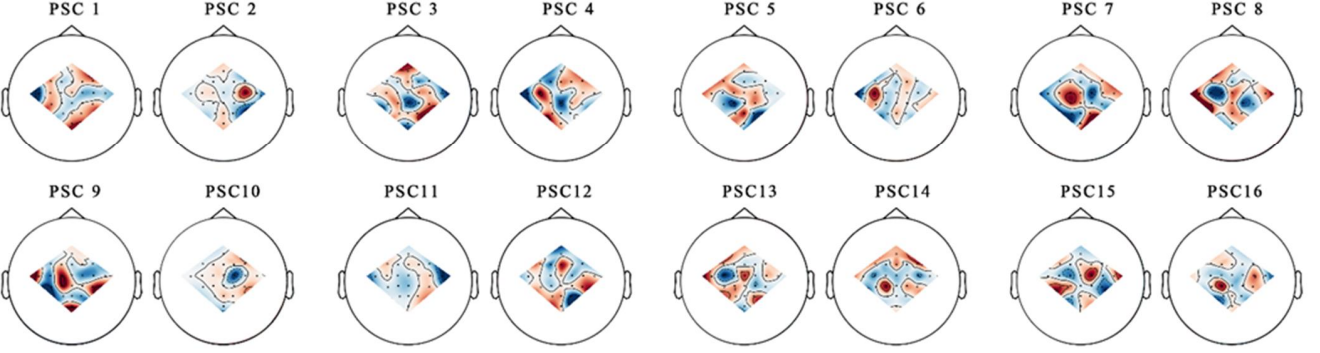

Fig. 5. The topographic maps of the learned spatial filters. Each PSC is obtained by a spatial filter. The black dots represent the electrode locations, from where the EEG signals were collected. The topographic map of each spatial filter is colored by the weight assigned for each electrode by the spatial filter. Red represents positive weight and blue represents negative. The deeper the color, the greater the magnitude of the weight.

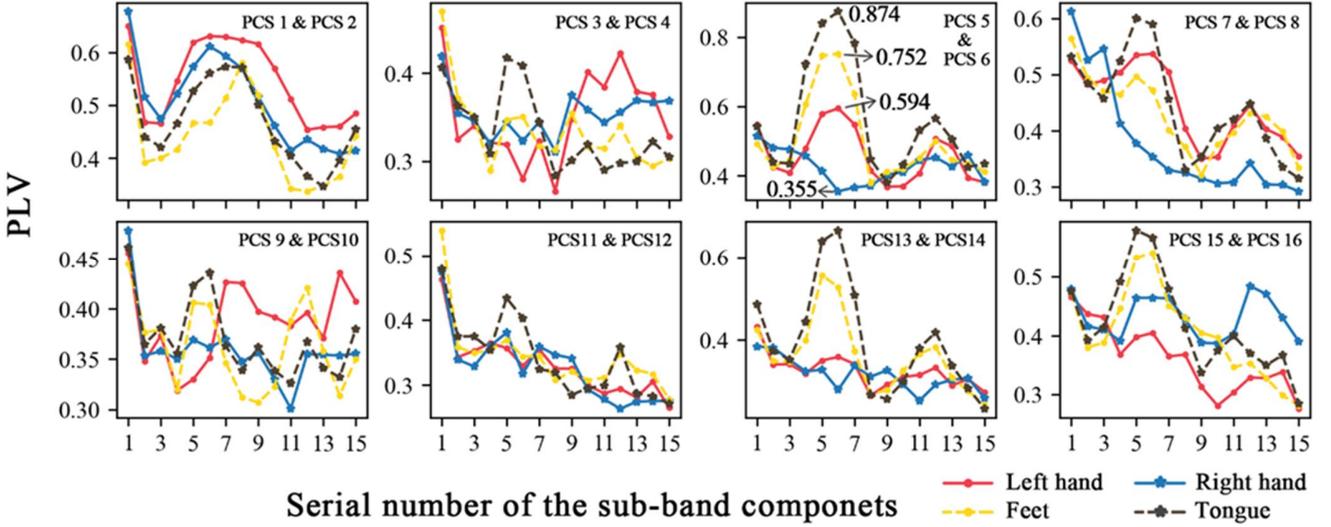

Fig. 6. PLVs of the extracted PSPs. Each PSP is formed by two sub-band components from the same frequency band of every two PSCs. The serial numbers of each two PSCs are indicated in the upper right corner of each subgraph. For instance, the PLVs of the PSPs composed of the sub-components of PSC 1 and PSC 2 are shown in the first subgraph. Both the PSCs consist of 15 sub-components, resulting in 15 PSPs. The x-axis represents the serial number of the 15 PSPs, while the y-axis represents PLV ranging from 0 to 1. The PLVs of the PSPs extracted from the samples of different classes are separately calculated, and the average PLV of samples from the same class is plotted.

can effectively mitigate estimation errors in the phase-to-amplitude transcoder.

As shown in Table II, PSNet demonstrates superior performance compared to other methods on MMIDB. PSNet outperforms DoNet on subjects S1, S3, S4, S5, and S8, but exhibits relatively lower performance on the remaining subjects.

### B. Phase Locking Value of PSP

The SpatialConv kernels of a trained PSNet on S1 are shown in Fig. 5. These kernels serve as optimal spatial filters for capturing the PSCs that reflect phase synchronization phenomena. Complex noise interference patterns make it challenging to exploit information solely by analyzing the topographic maps of these filters. Therefore, our focus shifts to analyzing the PLVs of the extracted PSPs, which offer a more intuitive reflection of the degree of phase synchrony.

In Fig. 6, we observed the PLVs of the potential PSPs formed by pairing the PSCs obtained through the filters illustrated in Fig. 5. For each PSP, the PLVs of the PSP extracted from samples of the same class are averaged, resulting in 4 average PLVs corresponding to left hand, right hand, feet, and tongue MI tasks. Notably, differences between the average PLVs of different MI samples can be observed in various subgraphs.

Particularly, in the third subgraph of Fig. 6, the PSPs comprised of the 6th sub-components of PSC 5 and PSC 6 exhibit extremely significant phase synchrony.

First, the average PLV of the PSPs from tongue MI samples is an impressive 0.874. The synchrony pattern of tongue motor imagery has received limited research attention, making this high PLV a groundbreaking discovery.

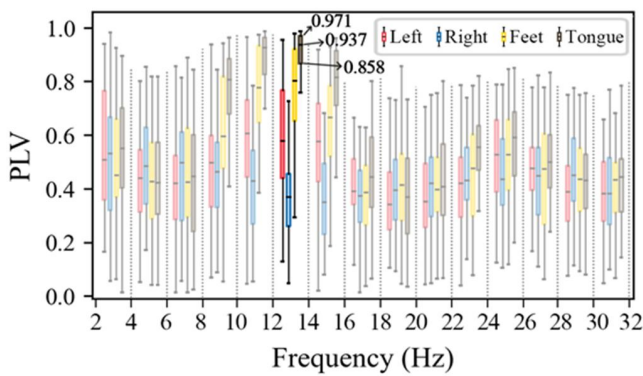

Fig. 7. Count the upper quartile, lower quartile, and median of the PLVs per class. The x-axis represents frequency, segmented into 15 frequency ranges by vertical dotted auxiliary lines. The 15 PSPs from PCS 5 and PCS 6 fall in the frequency ranges. There are 4 boxplots in each frequency range, corresponding to left hand, right hand, feet, and tongue from left to right. For aesthetic purposes, outliers were not shown.

Second, the average PLVs differ among samples from different classes, thereby enabling classification based on these differences.

Third, to further investigate, the subgraph is extended to a box graph as shown in Fig. 7. The 6th sub-components of PSC 5 and PSC 6 fall within 12-14Hz frequency range, which is highlighted in Fig. 7. It can be seen that the clustering degree of the PLVs for tongue MI samples is significantly higher compared to other samples. Specifically, three-fourths of the tongue MI samples have PLVs higher than 0.858, while half of them exceed 0.937. Surprisingly, one-fourth of the PLVs surpass 0.971, indicating near-complete phase locking with a value close to 1.

In summary, the PSP represent true synchrony patterns in the cortex in a physical sense, but not a mathematical fit result.

## V. DISCUSSIONS AND CONCLUSION

In the field of studying phase synchrony in brain activities, phase synchrony component self-organization scheme has been first proposed to automate noise suppression and channel selection, thereby reducing the need for manual intervention. A complete phase synchrony component self-organization pipeline involves the extraction of potential PSPs and the identification of their authenticity.

Based on this scheme, a DL network PSNet has been developed to implemented a concise phase synchrony component self-organization pipeline based on a classification task. The extraction of potential PSPs is achieved using convolution layers, while the authenticity of the potential PSPs is determined through classification accuracy. Experiment results have demonstrated that PSNet outperform existing state-of-the-art end-to-end MI classification networks.

The primary research objective of automatically identifying phase synchrony component pairs with high PLVs has been validated. Some of the PSPs extracted by a trained PSNet exhibit high PLVs for specific MI tasks. Additionally, differences in PLVs between different MI samples have been observed, enabling classification to be performed based on these variations.

Notably, in one subject, half of the tongue MI samples displayed PLVs higher than 0.937, indicating near-complete phase locking with a value close to 1. The synchrony pattern of tongue motor imagery has received limited attention in previous research, making this case a groundbreaking discovery.